# Generalization Of Classical Statistical Mechanics To Quantum Mechanics And Stable Property Of Condensed Matter

Y. C. Huang[1, 4, *]     F. C. Ma[2]     N. Zhang[3]

1. Theoretical Physics Group, Department of Applied Physics, Beijing University of Technology, Beijing 100022, P. R. China
2. Department of Physics, Liaoning University, Shenyang, 110036, P. R. China.
3. Optoelectronics Group, Cavendish Laboratory, Madingley Road, Cambridge CB3 0HE, U.K.
4. CCAST ( World Lab. ), P. O. Box 8730, Beijing, 100080, P. R. China

**Abstract**

Classical statistical average values are generally generalized to average values of quantum mechanics, it is discovered that quantum mechanics is direct generalization of classical statistical mechanics, and we generally deduce both a new general continuous eigenvalue equation and a general discrete eigenvalue equation in quantum mechanics, and discover that a eigenvalue of quantum mechanics is just an extreme value of an operator in possibility distribution, the eigenvalue $f$ is just classical observable quantity. A general classical statistical uncertain relation is further given, the general classical statistical uncertain relation is generally generalized to quantum uncertainty principle, the two lost conditions in classical uncertain relation and quantum uncertainty principle, respectively, are found. We generally expound the relations among uncertainty principle, singularity and condensed matter stability, discover that quantum uncertainty principle prevents from the appearance of singularity of the electromagnetic potential between nucleus and electrons, and give the failure conditions of quantum uncertainty principle. Finally, we discover that the classical limit of quantum mechanics is classical statistical mechanics, the classical statistical mechanics may further be degenerated to classical mechanics, and we discover that only saying that the classical limit of quantum mechanics is classical mechanics is mistake. As application examples, we deduce both Shrödinger equation and state superposition principle, deduce that there exist decoherent factor  from a general mathematical representation of state superposition principle, and the consistent difficulty between statistical interpretation of quantum mechanics and determinant property of classical mechanics is overcome.
**Key words**: quantum mechanics, uncertainty principle, classical statistics, operator, stability
PACS: 05.30.-d

## 1. Introduction

There are many difficult problems in quantum mechanics even today. People are quite familiar with classical mechanics, so to try grasping the relationship between classical and quantum mechanics is extremely important for profound understanding of quantum mechanics. Since quantum mechanics is statistical theory, it is natural to consider the corresponding relations between classical statistical mechanics and quantum mechanics. In fact, the both of them have statistical average values and uncertain relation, so we must do some research on how to generalize to quantum statistical average values and quantum uncertainty principle from classical statistical counterparts directly.

In the research about quantum and classical mechanics, Halliwell researched quantum-mechanical histories characterized by sequences of position samplings, constructed a probability distribution on the value of (discrete approximations to) the field equations, and classical correlations are exhibited [1], quantum-mechanical dualities from classical phase space is studied [2]. Ref. [3] investigated classical and quantum Nambu mechanics, the classical theory is reviewed and developed utilizing varied examples, and quantum theory is discussed in a parallel presentation and illustrated with detailed specific cases [3]. Superintegrability with third order invariants in quantum and classical mechanics is explored [4], the classical limit of quantum mechanics and the Fejer sum of the Fourier series expansion of a classical quantity are discussed [5], classical mechanics technique for quantum linear response is given [6], Ref.[7] researched the relation between quantum mechanical and classical parallel transport, the quantum BRST structure of classical mechanics is discussed [8], Anastopoulos studied information measures and classicality in quantum mechanics [9], Ref. [10] researched mixing quantum and classical mechanics, and classical and quantum mechanics of non-Abelian Chern-Simons particles are shown out [11], Ref. [12] further researchered intrinsic mechanism of entropy change in classical and quantum evolution.

---
[*] Corresponding author

Generally speaking, the sample space of classical statistical mechanics corresponds to the Hilbert space of quantum mechanics [13,14]. Quantum mechanics is the universal theory, while classical mechanics is just the limit situation of quantum mechanics when we ignore quantum effects, so it is natural to conclude that they have corresponding relation. But classical mechanics is a large scientific field, that quantum mechanics corresponds to which branch science is still a unsolved open problem, In classical mechanics, the moving state of a particle can be determined by Newton's Law and initial conditions. If the initial conditions are exactly fixed, then the solutions are exact. But if the initial conditions are given in the form of probability, especially initial conditions of microscopic particle are, in fact, the form of probability, then the solutions must also be in the form of probability.

## 2. Generalization of Classical Statistical Mechanics to Quantum Mechanics

Classical statistical mechanics can describe a system or a particle statistically. So in the system described by classical statistical mechanics, we not only give correspondence of classical statistical mechanics to quantum mechanics, but also give a general classical uncertain relation, and generalize the classical general uncertain relation to uncertainty principle of quantum mechanics, thus we can find that the classical uncertain relation has the exact connection with uncertainty principle of quantum mechanics.

For classical statistical mechanics, suppose $\rho = \rho(\vec{Y},t,t')$ $(\rho \geq 0)$, t' is initial time parameter, $\rho$ is a combination probability density function of t' and $\mathbf{Y}$ of the system at time $t$ [15].

At any time t, not losing generality, let's assume that a general function $f$ of a composite physical quantity in classical statistical mechanics is

$$f = f(A(\vec{Y},t,t'), B(\vec{Y},t,t')) \quad , \tag{1}$$

where A and B are two general functions, $\vec{Y}$ is a general vector variable, such as $\vec{Y} = \vec{R}$ ( that is the position vector of the particle). At time t, the average value of function $f$ is

$$\overline{f}(t) = \int_{-\infty}^{\infty} d\vec{Y} \int_{0}^{\infty} dt' \rho(\vec{Y},t,t') f(A(\vec{Y},t,t'), B(\vec{Y},t,t')) \quad , \tag{2}$$

thus we can have a general statistical average value in classical statistical mechanics as follows [15]

$$\overline{f}(t) = \int_{-\infty}^{\infty} d\vec{Y} \rho(\vec{Y},t) f(A(\vec{Y},t), B(\vec{Y},t)) \quad . \tag{3}$$

In order to discover the correspondence of classical statistical mechanics to quantum mechanics, we generalize $\rho(\vec{Y},t)$ from a general real function to a general representation form of complex function, thus, the natural generalization is that we can take $\rho(\vec{Y},t) = \Psi^*(\vec{Y},t)\Psi(\vec{Y},t)$, then $\rho(\vec{Y},t)$ as a whole quantity having the independence of phase, and in this case, $\rho(\vec{Y},t)$'s concrete function value doesn't change, only the form of $\rho(\vec{Y},t)$'s representation has changed, but, which just introduce new physics, i.e., quantum physics, then $\psi$ may be naturally moved to the right side of $f$, thus Eq.(3) may be generally and identically written as

$$\overline{f}(t) = \int_{-\infty}^{\infty} d\vec{Y} \Psi^*(\vec{Y},t) f(A(\vec{Y},t), B(\vec{Y},t)) \Psi(\vec{Y},t) \quad . \tag{4}$$

When $f$ is an operator function, that is, $f(A,B)$ is changed to $\hat{f}(\hat{A},\hat{B})$, the average value of $\hat{f}$, i.e., the average value in quantum mechanics, is

$$\overline{\hat{f}(t)} = \int_{-\infty}^{\infty} d\vec{Y} \Psi^*(\vec{Y},t) \hat{f}(\hat{A}(\vec{Y},t), \hat{B}(\vec{Y},t)) \Psi(\vec{Y},t) \quad , \tag{5}$$

in which $\vec{Y}$ can be coordinate, momentum and so on. Therefore, a general statistical average value in classical statistical mechanics is generalized to a general average value in quantum mechanics.

We now generally deduce eigenvalue equation of quantum mechanics. Because, in terms of operator algebra [16], $\Psi(\vec{Y},t)$ and $\Psi^*(\vec{Y},t)$ are linearly independent, we make the variation of the difference of quantum average value (5) and classical average value (4) about functions $\Psi(\vec{Y},t)$ and $\Psi^*(\vec{Y},t)$, we obtain

$$\delta A = \delta(\overline{\hat{f}(t)} - \overline{f}(t)) = \int_{-\infty}^{\infty} d\vec{Y} \{\delta\Psi^*(\hat{f}-f)\Psi + [(\hat{f}-f)\Psi]^*\delta\Psi\} \quad , \tag{6}$$

where $\hat{f}$ is Hermitian operator. Using linear independence of functions $\Psi(\vec{Y},t)$ and $\Psi^*(\vec{Y},t)$, we deduce

$$(\hat{f}-f)\Psi = 0 \quad , \tag{7}$$

Eq.(7) is a general eigenvalue equation whose eigenvalue corresponding to classical statistical mechanics may be a general function of continuous variable, then Eq.(7) is the most general and the new eigenvalue equation, and using Eq.(7) we can easily get the all known eigenvalue equations, e.g., $f$ may be a constant eigenvalue, see Section 4. Therefore, we discover that eigenvalue of quantum mechanics is just extreme value of operator in possibility distribution, and the eigenvalue $f$ is just classical observable quantity. Accordingly, the general relation between observable quantity $f$ of classical statistical mechanics and operator $\hat{f}$ of quantum mechanics is given, and we naturally obtain that the eigenvalue of operator in quantum mechanics is real number ( in classical statistical mechanics ).

Now we give a general classical uncertain relation, and then generalize it to a general quantum uncertainty principle.

Suppose $\Delta A(\vec{Y},t) = A(\vec{Y},t) - \overline{A}$, $\Delta B(\vec{Y},t) = B(\vec{Y},t) - \overline{B}$. If $\xi$ is real, suppose $f = (\Delta A \xi + \Delta B)^2$ in Eq.(3), then

$$\overline{(\Delta A \xi + \Delta B)^2} = \int_{-\infty}^{\infty} d\vec{Y} \rho(\vec{Y},t)(\Delta A(\vec{Y},t)\xi + \Delta B(\vec{Y},t))^2 \quad , \tag{8}$$

or

$$G = \overline{\Delta A^2}\xi^2 + 2\overline{\Delta A \Delta B}\xi + \overline{\Delta B^2} \geq 0 \quad . \tag{9}$$

Then we get that the condition of keeping Eq.(9) to stand is

$$\overline{\Delta A^2} \cdot \overline{\Delta B^2} \geq (\overline{\Delta A \Delta B})^2 \quad , \tag{10}$$

Eq.(10) is just a general uncertain relation of classical statistical mechanics. We now generalize the above research to a general quantum uncertainty principle.

Suppose $f = (\Delta A \xi + \Delta B)^2$ in equation (4), and generalize it to the representation form of complex function, then the best natural generalization is $f = (\Delta A \xi + \Delta B)^*(\Delta A \xi + \Delta B)$. Then $\Delta A \xi + \Delta B$ can generally be complex. Thus $f$ as a whole quantity has the independence of phase, and the value of $f$ is not changed, just the representation form is changed.

Accordingly, suppose $\Delta A \to \Delta \hat{A}, \ \Delta B \to \Delta \hat{B}, \ \xi \to i\zeta, \zeta > 0$, then we deduce $\hat{f} = \left|\Delta \hat{A}\zeta - i\Delta \hat{B}\right|^2$,

thus, we obtain

$$\overline{\left|\Delta \hat{A}\zeta - i\Delta \hat{B}\right|^2} = \int \Psi^* \left|\Delta \hat{A}\zeta - i\Delta \hat{B}\right|^2 \Psi d\vec{Y} \geq 0 \quad . \tag{11}$$

When

$$\left[\hat{A}, \hat{B}\right] = i\hat{C} \quad , \tag{12}$$

Eq. (11) is expressed as

$$F = \overline{\Delta \hat{A}^2} \zeta^2 + \overline{\hat{C}}\zeta + \overline{\Delta \hat{B}^2} \geq 0 \quad , \tag{13}$$

Eqs. (9) and (13) are important one-to-one formulas when we want to find general uncertain relations in classical statistical mechanics and quantum mechanics, respectively.

It is well known that many textbooks and articles think of that the condition of keeping equation (13) to stand is

$$\overline{\Delta \hat{A}^2} \cdot \overline{\Delta \hat{B}^2} \geq \frac{\overline{\hat{C}}^2}{4} \quad , \tag{14}$$

i.e., Eq.(14) is the uncertainty principle in quantum mechanics. But requesting equation (10) and (14) only are not sufficient, because when using Eqs.(9) and (13) we have

$$G = \overline{\Delta A^2}(\xi + \frac{\overline{\Delta A \Delta B}}{\overline{A^2}})^2 + \overline{\Delta B^2} - \frac{(\overline{\Delta A \Delta B})^2}{\overline{\Delta A^2}} \geq 0 \quad , \tag{15}$$

$$F = \overline{\Delta \hat{A}^2}(\zeta + \frac{\overline{\hat{C}}}{4\overline{\Delta \hat{A}^2}})^2 + \overline{\Delta \hat{B}^2} - \frac{(\overline{\hat{C}})^2}{4\overline{\Delta \hat{A}^2}} \geq 0 \quad . \tag{16}$$

So only when $\overline{\Delta A^2} \neq 0$ & $\overline{\Delta B^2} \neq 0$ and $\overline{\Delta \hat{A}^2} \neq 0$ & $\overline{\Delta \hat{B}^2} \neq 0$, we can derive inequalities (10) and (14) from (15) and (16), respectively. That is to say that only giving conditions of equations (10) and (14) are insufficient, because which omit the very important conditions of $\overline{\Delta A^2} \neq 0$ & $\overline{\Delta B^2} \neq 0$ and $\overline{\Delta \hat{A}^2} \neq 0$ & $\overline{\Delta \hat{B}^2} \neq 0$. And because when $(\overline{\Delta A \Delta B})^2$ and $\overline{\hat{C}}^2/4$ are not zero, we must have conditions $\overline{\Delta A^2} \neq 0$ & $\overline{\Delta B^2} \neq 0$ and $\overline{\Delta \hat{A}^2} \neq 0$ & $\overline{\Delta \hat{B}^2} \neq 0$, otherwise we cannot have inequalities (10) and (14). In particular, if the right sides of inequalities (10) and (14) are zero, the result is absurd, because that zero times any value must equals to zero, but which say also the result of the multiplication is greater than a positive number not to equal to zero. Thus, it is wrong that some references and books say, when we discuss uncertainty principle, there may be $\overline{\Delta \hat{A}^2} = 0$. $\overline{\Delta \hat{A}^2}$ and/or $\overline{\Delta \hat{B}^2}$ can only approach zero, but cannot be equal to zero, which are seriously distinct in mathematics.

Therefore, we not only generalize classical statistical mechanics to quantum mechanics, but also extend a classical general statistical uncertain relation to quantum uncertainty principle.

### 3. Application of Quantum Uncertainty Principle to Stability of Condensed Matter

According to classical electromagnetism, nucleuses are surrounded by rotating electrons. Because of rotation acceleration, electrons emit energy and lose kinetic energy, so they will fall into the nucleus in the end. But why don't the electrons fall into the nucleus in the end ? And they are stable in nature. That is because in this situation, quantum mechanics must have zero point energy and corresponding minimum average radius, and quantum uncertainty principle gives that the smaller $\overline{\Delta \hat{A}^2} = \overline{\Delta \hat{x}^2}$ is, the larger $\overline{\Delta \hat{B}^2} = \overline{\Delta \hat{p}^2}$ is, and $\overline{\Delta \hat{x}^2} \neq 0$. Just because of $\overline{\Delta \hat{A}^2} = \overline{\Delta \hat{x}^2} \neq 0$, $\overline{\Delta \hat{B}^2} = \overline{\Delta \hat{p}^2} \neq 0$ and equation (16), we can derive quantum uncertainty principle. When we substitute an approximate quantum uncertain relation to corresponding Hamiltonian and calculate extreme value, then zero point energy and corresponding minimum average are easily derived. Thus, the radius of ground state is determined mainly by quantum uncertainty principle (such as H, He$^+$, Li$^{++}$). Quantum uncertainty principle not only deters electrons to fall into the nucleus but also mainly maintain zero point energy.

In fact, when we use other methods to deduce quantum uncertainty principle, we also need $\overline{\Delta \hat{A}^2} \neq 0$ and $\overline{\Delta \hat{B}^2} \neq 0$, even in the deduction process of Schwarz inequality [17,18]. Furthermore, when we make $[\hat{A}, \hat{B}] = i\hat{C}$ and Schwarz inequality together to deduce quantum uncertainty principle, it is still not allow $\overline{\Delta \hat{A}^2} = 0$ and/or $\overline{\Delta \hat{B}^2} = 0$, otherwise the absurd result that zero multiplies any value equals to zero but also greater than a positive number not being zero will be present. From the above research we know that all of the approaches used to deduce quantum uncertainty principle are equivalent. So when we refer to quantum uncertainty principle, we must include $\overline{\Delta \hat{A}^2} \neq 0$ and $\overline{\Delta \hat{B}^2} \neq 0$, or else it is inexact and insufficient. All of the books and articles relative to quantum uncertainty principle should keep noticing these.

After deducing the minimum average radius, plus the action of Pauli exclusive principle, a stable atom can be formed, because the other electrons of a chemical element can be filled one by one in order from the electrons of the most inside layer to the electrons of higher energy level, consequently the stable size of construction of the atomic electron layer can be formed.

When is quantum uncertainty principle ineffective? In the extreme situation when gravitation is so strong that it meets the criterion of neutron star, because, at this time, the ground state radius maintained by quantum uncertainty principle collapses, and electrons fall into the nucleus to form neutrons together with protons, so, at this time, both $\overline{\Delta \hat{A}^2} \neq 0$ and inequality (14) do not come into existence. Thus, it is discovered that quantum uncertainty principle prevents from the appearance of singularity of the electromagnetic potential between nucleus and electrons, therefore, the zero point energy ( of the most inner layer electron ) that makes atoms stable is not only different from gravitation, strong, week and electromagnetic interaction energy but also different from the energy of boson condensation attractive force and Fermi exclusive force. Condensed matter stability is built on the stability of ground state of quantum electromagnetic interaction in atoms, while the stability of the ground state is determined mainly by quantum uncertainty principle. If the ground state collapses, all the other states transit to the ground state and collapse. In the end, the condensed matter world collapses rapidly, which is because almost all condensed matter is coagulated together by quantum electromagnetism interaction, and the condensed matter world is very rich, even the interaction of the all chemistry varieties is just quantum electromagnetism interaction.

### 4. Classical Correspondence of Quantum Mechanics to Classical Statistical Mechanics

In classical statistical mechanics, for discrete quantities $f_i$, there is a set $\{\rho_i(\vec{Y},t)\}$ ($\rho_i \geq 0$) of classical statistical sample space [15], $\rho_i(\vec{Y},t)$ represents the possibility distribution that $f$ takes $f_i$. For all systems consisted of $\rho_i(\vec{Y},t)$ and $f_i$, $i \in Z$, there is a general average relation of the assemblage consisted of all $\rho_i(\vec{Y},t)$ and $f_i$ as follows

$$\overline{f}(t) = \int_{-\infty}^{\infty} d\vec{Y} \sum_i a_i(t) \rho_i(\vec{Y},t) f_i \ . \tag{17}$$

where $\sum_i a_i = 1, a_i \geq 0,$ then $a_i$ represents the probability taking the system of $\rho_i(\vec{Y},t)$ and $f_i$.

As the discussion above, for any i, there may exist a similar relation

$$\rho_i(\vec{Y},t) = \phi^*_i(\vec{Y},t) \phi_i(\vec{Y},t), \tag{18}$$

and we can define $a_i(t) = |c_i(t)|^2$, thus, we have

$$\overline{f}(t) = \int_{-\infty}^{\infty} d\vec{Y} \sum_i c_i^*(t) \phi_i^*(\vec{Y},t) f_i c_i(t) \phi_i(\vec{Y},t) \ . \tag{19}$$

In quantum mechanics, there is a general relation $\Psi(\vec{Y},t) = \sum_i c_i(t)\phi_i(\vec{Y},t)$, then Eq.(5) can be rewritten as

$$\overline{\hat{f}(t)} = \int_{-\infty}^{\infty} d\vec{Y} \left( \sum_{i,j} c^*_i(t) c_j(t) \phi^*_i(\vec{Y},t) \hat{f} \phi_j(\vec{Y},t) \right) . \quad (20)$$

We now generally deduce discrete eigenvalue equation of quantum mechanics. Because, in terms of operator algebra [17], $\phi_i(\vec{Y},t)$ and $\phi^*_i(\vec{Y},t)$ are linearly independent, we make the variation of the difference of quantum average value (20) and classical average value (19) about linear independent functions $\phi_i(\vec{Y},t)$ and $\phi^*_i(\vec{Y},t)$, we obtain

$$\delta A = \delta(\overline{\hat{f}(t)} - \overline{f(t)})$$

$$= \int_{-\infty}^{\infty} d\vec{Y} \sum_{i,j} c^*_i(t) c_j(t) \{\delta\phi^*_i(\vec{Y},t)(\hat{f} - \delta_{ij} f_i)\phi_j(\vec{Y},t) + [(\hat{f} - \delta_{ij} f_i)\phi_i(\vec{Y},t)]^* \delta\phi_j(\vec{Y},t)\}, \quad (21)$$

because $\hat{f}$ is Hermitian operator. Using linear independence of functions $\phi_i(\vec{Y},t)$ and $\phi^*_i(\vec{Y},t)$, we deduce

$$\sum_j c^*_i(t) c_j(t)(\hat{f} - \delta_{ij} f_i)\phi_j(\vec{Y},t) = 0, \text{ or } \sum_i c_j(t) c^*_i(t)[(\hat{f} - \delta_{ij} f_i)\phi_i(\vec{Y},t)]^* = 0, \quad (22)$$

When $c^*_i(t), c_j(t) \neq 0$, we obtain a general eigenvalue equation

$$\hat{f}\Psi(\vec{Y},t) = \sum_j c_j(t) f_j \phi_j(\vec{Y},t), \text{ or } \hat{f}\phi_j(\vec{Y},t) = f_j \phi_j(\vec{Y},t), \quad (23)$$

where we have used the linear independence of $\phi_j(\vec{Y},t), \ j \in Z$. Eq.(23) is a general discrete eigenvalue equation. Therefore, we discover that eigenvalue of quantum mechanics is just extreme value of operator in possibility distribution, and the eigenvalues $f_j, j \in Z$ are just classical observable quantities. Accordingly, the general relation between observable quantities $f_j, j \in Z$ of classical statistical mechanics and operator $\hat{f}$ of quantum mechanics is given, and we naturally obtain that the eigenvalue of operator in quantum mechanics is real number ( in classical statistical mechanics ). Consequently, we cancel a basic presumption of operator, eigenvalue and measuring probability of quantum mechanics by using this letter's results.

On the other hand, Eq.(20) can be rewritten as

$$\overline{\hat{f}(t)} = \int_{-\infty}^{\infty} d\vec{Y} \left( \sum_i c^*_i c_i \phi^*_i \hat{f} \phi_i + \sum_{i,j} c^*_i c_j \phi^*_i \hat{f} \phi_j \right), (i \neq j) \quad , \quad (24)$$

When $\int_{-\infty}^{\infty} d\vec{Y} \left( \phi^*_i \hat{f} \phi_j \right) = 0 \ (i \neq j)$, the average value in quantum mechanics has the same form as the average value (17) in classical statistical mechanics but $\hat{f}$ being operator, and when $\hat{f} \to f$ is a classical correspondence function, the average value in quantum mechanics is degenerated to the average value (17) in classical statistical mechanics, therefore, we discover that, in general case, the average value in quantum mechanics is different to the average value in classical statistical mechanics. Thus, for any operator $\hat{f}$ ( even a constant number ), there is $\int_{-\infty}^{\infty} d\vec{Y} \left( \phi^*_i \hat{f} \phi_j \right) = 0$ $(i \neq j)$, then the quantum system is degenerated ( or call, decohered ) to classical statistical mechanics, because, in classical statistical mechanics, there are not the contribution from the intercrossing terms. The decoherent factor of (i,j) part is $\alpha_{ij}(\vec{Y},t) = \phi^*_i(\vec{Y},t)\phi_j(\vec{Y},t)$ $(i \neq j)$, when $\alpha_{ij}(\vec{Y},t) \to 0$ $(i \neq j)$, the quantum system is decoherent. Thus, in quantum mechanics, the condition of decohering to classical statistical mechanics is given. Further when the corresponding physical quantity satisfies the relation $f \gg \hbar$, then the statistical effect can be neglected, i.e., $\vec{Y}$ of $f$ in Eq.(3) has nothing to do with random variable $\vec{Y}$ in $\rho(\vec{Y},t)$ and $d\vec{Y}$ of Eq.(3), we can denote $\vec{Y}$ in $f$ as $\vec{X}$, accordingly, the classical statistical mechanics is degenerated to classical mechanics, because, at this time, it follows from the above discussion that

$$\overline{f}(t,\vec{X}) = f(t,\vec{X}), \quad (\int_{-\infty}^{\infty} d\vec{Y} \rho(\vec{Y},t) = 1), \tag{25}$$

Eq.(25) is meaningful, because which means that mechanical quantity $f$ has determinative value at every $\vec{X}$, and has nothing to do with random variable $\vec{Y}$. Therefore, we discover that the classical limit of quantum mechanics is classical statistical mechanics, and the classical statistical mechanics may further be degenerated to classical mechanics, and we prove that the theory that only thinks of quantum mechanics may directly degenerate to classical mechanics is mistake, because quantum mechanics is a theory of statistical property, then its classical correspondence should be a theory of classical statistical property, but classical mechanics of a determinative property.

As application examples, using eigenvalue equation (7) or (23), we now deduce both Shrödinger equation and state superposition principle. For plane wave $\Psi = ce^{i(\vec{p}\cdot\vec{r}-Et)/\hbar}$, using Eq.(7) we obtain operators $\hat{\vec{p}} = -i\hbar\nabla$ and $\hat{E} = i\hbar\partial/\partial t$ of eigenvalues momentum $\vec{p}$ and energy E, respectively. On the other hand, eigenvalue energy $E = T(\vec{p}) + V(\vec{r})$, then we obtain Hamiltonian operator $\hat{H} = \hat{T}(-i\hbar\nabla) + \hat{V}(\hat{\vec{r}})$, because the other wave functions can be expanded by means of Fourier plane wave, thus we achieve a general Shrödinger equation $i\hbar \dfrac{\partial}{\partial t}\Psi = \hat{H}\Psi$.

For any $i \in Z$, there is $i\hbar\dfrac{\partial}{\partial t}\Psi_i = \hat{H}\Psi_i$, we now can deduce a general state superposition principle. Because Shrödinger equation is linear, using superposition property of linear equation, for any complex constant $\alpha_i$, we can get $i\hbar \frac{\partial}{\partial t}\Psi = \hat{H}\Psi$, where $\Psi = \sum_i \alpha_i \Psi_i$, $i \in Z$, i.e., $\Psi = \sum_i \alpha_i \Psi_i$ is a general mathematical representation of state superposition principle. Accordingly, when putting the general mathematical representation $\Psi = \sum_i \alpha_i \Psi_i$ of state superposition principle into Eq.(5) we can deduce that there is decoherent factor from the general mathematical representation of state superposition principle, and the hard consistent difficulty between statistical interpretation of quantum mechanics and determinant property of classical mechanics is overcome in this paper.

The researches above in this letter satisfy the quantitative causal principle ( e.g., changes ( cause ) of some quantities in Eq.(7) must cause relative changes ( result ) of the other quantities in Eq.(7) so that Eq.(7)'s right side keeps no-loss-no-gain [19], i.e., maintains zero ) and are not only consistent but also in accordance with the actual known physics.

## 5. Summary and Conclusion

The combination probability density function of classical statistical mechanics is given, and it has been broadened to a general representation of complex function. Classical statistical average values are generally generalized to average values of quantum mechanics, it is discovered that classical statistical mechanics may be generalized to quantum mechanics, this letter generally deduces both a new general eigenvalue equation of continuous variable and a general discrete eigenvalue equation in quantum mechanics, and discovers that eigenvalue of quantum mechanics is just extreme value of operator in possibility distribution, the eigenvalue $f$ is just classical observable quantity, it is naturally obtained that the eigenvalue of operator in quantum mechanics is real number ( in classical statistical mechanics ), and we discover that, in general case, the average value in quantum mechanics is different to the average value in classical statistical mechanics, in quantum mechanics, the condition of decohering to classical statistical mechanics is given. A general classical statistical uncertain relation is further given, the general classical statistical uncertain relation is generally generalized to quantum uncertainty principle, the two lost conditions in classical uncertain relation and quantum uncertainty principle, respectively, are found. This letter generally expounds the relations among uncertainty principle, singularity and condensed matter stability, it is discovered that quantum uncertainty principle prevents from the appearance of singularity of the electromagnetic potential between nucleus and electrons, and this letter gives the failure conditions of quantum uncertainty principle. Finally, this letter discovers that the classical limit of quantum mechanics is classical statistical mechanics, the classical statistical mechanics may further be degenerated to classical mechanics, and it is proved that the theory that only thinks of quantum mechanics may directly degenerate to classical mechanics is mistake, because quantum mechanics is a theory of statistical property, then its classical correspondence should be classical statistical mechanics (a theory of classical statistical property ), but classical mechanics (of determinative property ). As application examples, using the deductive eigenvalue equation, we deduce both Shrödinger equation and state superposition principle, we deduce that there is decoherent factor from a general mathematical representation of state superposition principle, in final, the consistent difficulty between statistical interpretation of quantum mechanics and determinant property of classical mechanics is overcome in this paper.
.